\documentclass{article}

\usepackage{arxiv}

\usepackage[utf8]{inputenc} 
\usepackage{hyperref}       
\usepackage{url}            
\usepackage{tikz}

\title{Evaluating Non-aligned Musical Score Transcriptions with MV2H}

\author{
  Andrew McLeod\\
  Graduate School of Informatics\\
  Kyoto University\\
  \texttt{mcleod@sap.ist.i.kyoto-u.ac.jp}
}

\begin{document}
\maketitle

\begin{abstract}
The original MV2H metric \cite{McLeod:18a} was designed to evaluate systems which transcribe from an input audio (or MIDI) piece to a complete musical score. However, it requires both the transcribed score and the ground truth score to be time-aligned with the input. Some recent work has begun to transcribe directly from an audio signal into a musical score, skipping the alignment step. This paper introduces an automatic alignment method based on dynamic time warp which allows for MV2H to be used to evaluate such non-aligned transcriptions. This has the additional benefit of allowing non-aligned musical scores---which are significantly more widely available than aligned ones---to be used as ground truth. The code for the improved MV2H, which now also includes a MusicXML parser, and allows for key and time signature changes, is available at \url{www.github.com/apmcleod/MV2H}.
\end{abstract}

\section{Introduction}
MV2H (for \textbf{M}ulti-pitch, \textbf{V}oice, \textbf{M}eter, note \textbf{V}alue, and \textbf{H}armony) \cite{McLeod:18a} was introduced to evaluate ``complete'' automatic music transcription (AMT) systems which transcribe an input audio (or MIDI) piece into a musical score, complete with voice (or instrument) separation, a time signature and metrical structure, note values, and harmonic information such as a key signature and chord progression. Its design was based on the principal of \textit{disjoint penalties}: that a single transcription error should only be penalized once, even if that error causes multiple mistakes (for example, an error in time signature also causing note value errors).

Traditionally, multi-pitch detection systems (arguably the most important step for an AMT system) have aligned their output in time with the input musical piece (e.g., \cite{sigtia2016end,Kelz2016,hawthorne2018onsets}), and while complete AMT systems are uncommon, the design of MV2H relied on such an alignment being present both for the transcription and the ground truth. However, recently a few complete AMT systems have been proposed which skip this alignment and instead output a musical score (or at least a text-based representation of a musical score) directly in an end-to-end fashion \cite{carvalho2017towards,romanend}, making them impossible to evaluate with MV2H.

For these methods, Cogliati and Duan \cite{Cogliati2017} describe a metric for evaluating complete transcriptions of non-aligned musical scores, using dynamic time warping (DTW; \cite{sakoe1990dynamic}) for alignment. Their metric, however, does not share our principal of disjoint penalties. Nakamura et al. \cite{nakamura2018towards} describe a method to evaluate multi-pitch detection and rhythmic aspects of a non-aligned transcription, relying on an existing method for alignment \cite{nakamura2017performance}.

This paper describes a simple alignment algorithm based on DTW to allow for such non-aligned transcriptions to be evaluated with MV2H. This also enables digital musical scores (such as MusicXML files \cite{good2001musicxml}, for which a parser is included) to be used as the ground truth without requiring any additional manual alignment or annotation. Furthermore, we extend MV2H to be able to evaluate transcriptions and ground truths which may contain key and time signature changes.

\section{MV2H Improvements}
\subsection{Automatic Alignment}
To perform an alignment between two musical scores, we first consider each musical score (transcribed and ground truth) as a sequence of chords, where each chord is a set of the notes which share an onset position. These sequences are then aligned using DTW, where the distance between two chords is defined as the F-measure between its notes, regarding only pitch (not value): a matched note is a true positive, an unmatched transcribed note is a false positive, and an unmatched ground truth note is a false negative. When a chord contains two notes of identical pitch (for example, in different instruments), each must match with a different note to count as a true positive.

We set the penalty for an insertion or deletion to $0.6$. This penalty makes the algorithm align chords which share no notes (distance of $1.0$ per chord pair) rather than have a series of insertions and deletions (distance of $1.2$ per chord pair), ensuring a more linear alignment. However, it is small enough that the distance between two chords dominates this penalty, ensuring that two chords that share even a single note are still aligned if possible.

After running this DTW, we treat aligned chords as anchor points, and set the relative timing for non-aligned musical features (chords, as well as bars, beats, sub beats, key signatures, chord symbols, etc.) based on the proportional distance between the surrounding anchor points, assuming constant tempo. The relative timing of musical features which lie after the final anchor point or before the first anchor point are set based on the preceding or succeeding anchored section, respectively. Once two musical scores are aligned, the thresholds for matching their note onsets, durations, and metrical groupings are all set to $0 ms$ (such that they must align exactly to count as a match).

\subsection{Key Changes}
MV2H uses the standard key detection evaluation, used by both mir\_eval \cite{Raffel2014} and the Music Information Retrieval Evaluation Exchange (MIREX) \cite{Mirexkey}. This assigns a score of $1.0$ to the correct key, $0.5$ to a key which is a perfect fifth too high or low, $0.3$ to the relative major or minor of the correct key, $0.2$ to the parallel major or minor of the correct key, and $0.0$ otherwise.

The new version, which allows key changes, evaluates each \textit{continuous key section} (the section of a piece which contains no time signature changes in the transcription and the ground truth) with the standard metric. The final key evaluation for the new version of MV2H is calculated as the sum of all of those scores, weighted by the proportion of the piece for which the given score is assigned. This is illustrated in Figure \ref{fig:key} for an example transcription.

\begin{figure}
    \centering
    \begin{tikzpicture}
    \draw [color=red, dashed] (0,1.5) -- (0,-1.3);
    \draw [color=red, dashed] (1,1.5) -- (1,-1.3);
    \draw [color=red, dashed] (2,1.5) -- (2,-1.3);
    \draw [color=red, dashed] (4,1.5) -- (4,-1.3);
    \node [anchor=east] at (0,1) {Ground Truth:};
    \draw [|-|,line width=0.5mm] (0,1) -- node[above] {D maj} (2,1);
    \draw [-|,line width=0.5mm] (2,1) -- node[above] {G min} (4,1);
    \node [anchor=east] at (0,0) {Transcription:};
    \draw [|-|,line width=0.5mm] (0,0) -- node[above] {D maj} (1,0);
    \draw [-|,line width=0.5mm] (1,0) -- node[above] {G maj} (4,0);
    \node [anchor=east] at (0,-1) {Per section:};
    \node at (0.5,-1) {$1.0$};
    \node at (1.5,-1) {$0.5$};
    \node at (3,-1) {$0.2$};
    \node at (1,-1.8) {Overall score $= (0.25)(1.0) + (0.25)(0.5) + (0.5)(0.2) = \textbf{0.475}$};
    \end{tikzpicture}
    \caption{The new key detection evaluation on an example transcription of a piece with key changes. Continuous key sections are separated by the vertical dashed lines, and the evaluation of each section is labeled as ``Per section''. The overall score is $0.475$ and its calculation is written as a sum of (proportion)(score) products.}
    \label{fig:key}
\end{figure}

\subsection{Time Signature Changes}
The metrical structure of a transcription is evaluated in MV2H using metrical F-measure. The metric is based on groupings at the bar, beat, and sub-beat level (because these three levels exactly define a time signature), where a grouping is represented by its start and end time. Transcribed groupings are compared to ground truth groupings, and any whose start and end times are both within $50 ms$ of each other, regardless of level, are counted as true positives. Unmatched transcribed groupings count as false positives, and unmatched ground truth groupings count as false negatives. The final metric is the standard F-measure using those counts.

The new metrical F-measure, which allows time signature changes, is quite similar, based on the same groupings (which are well-defined, even across changes in time signature). However, when evaluating an automatically-aligned transcription and ground truth, two groupings count as a match only if their start and end points are exactly aligned. The rest of the calculation is performed identically.

\section{Conclusion}
This paper has introduced an automatic alignment method to allow MV2H (for \textbf{M}ulti-pitch, \textbf{V}oice, \textbf{M}eter, note \textbf{V}alue, and \textbf{H}armony) \cite{McLeod:18a} to be used to evaluate non-aligned transcriptions and ground truth musical scores. This has two benefits: (1) it allows for the evaluation of end-to-end systems which do not produce such an alignment, and (2) it allows for (much more widely available) non-aligned musical scores (e.g., MusicXML, for which a parser is included) to be used as ground truth. Furthermore, the metric is now able to evaluate transcriptions and ground truths which may contain key and time signature changes. The described improvements vastly increase the types of transcription systems which can use MV2H for evaluation. The code is available at \url{www.github.com/apmcleod/MV2H}.

\section{Acknowledgements}
Thanks to Eita Nakamura for providing the MusicXML parser.

\bibliographystyle{unsrt}  
\bibliography{references}

\end{document}